\documentclass[twocolumn,showpacs]{revtex4}
\usepackage{epsfig}

\begin{document}

\title{Ferromagnetism and metallic state in digital (Ga,Mn)As heterostructures} 

\author{Stefano~Sanvito}
\email{e-mail: sanvitos@tcd.ie} 
\affiliation{Physics Department, Trinity College, Dublin 2, Ireland}

\date{\today}

\begin{abstract}
We present an extensive density functional theory study of the 
electronic, magnetic and transport properties of 
GaAs and AlAs digital ferromagnetic heterostructures. These can be obtained by 
$\delta$-doping with Mn the GaAs layers of a GaAs/AlAs superlattice.
Our analysis spans a range of Mn concentrations and considers the
presence of compensating defects such as As antisites. 

In the defect-free case all the heterostructures studied present an half-metallic 
electronic structure.
In contrast when As antisites are present the half-metallic state is destroyed 
and the heterostructures behave as dirty planar metals. In this case they show a large 
$p$-type metallic conductance in the Mn plane mainly due to majority spin electrons, 
and an $n$-type hopping-like conductance in the GaAs planes mainly due to minority
spin electrons. This suggests that if the As antisites can be kept far from the Mn
planes, spatial separation of the different spin currents can be achieved.
Finally we show that in the case of AlAs/(Ga,Mn)As digital ferromagnetic
heterostructures the AlAs/GaAs valence band offset produces an additional confining
potential for the holes responsible for the ferromagnetism. Therefore the ferromagnetic
coupling between the Mn ions becomes larger and more robust to the presence
of As antisites.

\end{abstract}

\maketitle

\section{Introduction}

(Ga,Mn)As \cite{Ohno1} is the prototype of a new class of magnetic materials 
named diluted magnetic semiconductors (DMS) \cite{Ohno2,Ohno3}. These are obtained by
doping ordinary semiconductors with transition metals. In the case of (Ga,Mn)As
the Mn ions occupy the Ga sites and provide both localized spins ($S$=5/2) and holes. 
The spin holes are then antiferromagnetically coupled to the Mn ions and this 
gives rise to hole-mediated long range ferromagnetism
via a Zener-like mechanism \cite{Dietl}.

The potential impact of this material on the semiconductor industry is huge, since
it opens the possibility of adding ferromagnetism to the AlAs/GaAs system, an
important step toward the implementation of the spin degree of freedom in an
electronic device \cite{Prinz}. To date several concept devices have been 
demonstrated, including spin polarized light emitters \cite{Aws1} and electrically 
controlled ferromagnetism \cite{Aws2}, and one can envision (Ga,Mn)As among the
building blocks for a spin-based quantum computer scheme \cite{DDV}.

Despite these indisputable successes the critical temperature ($T_C$) of
(Ga,Mn)As hardly exceeds 110~K \cite{Ohno2,Ohno3} and this poses severe
limitations to future commercial applications. 
At present low temperature annealing processing has allowed an increase of
the Curie temperature from 110~K to about 170~K \cite{not,nit}, but room 
temperature ferromagnetic (Ga,Mn)As appears difficult to produce.
However these recent annealing experiments have demonstrated an important point: 
the low critical temperature of (Ga,Mn)As is not an intrinsic limitation of
the material, but is affected by the inability to control the Mn and holes
concentration independently. 

A possible way to improve the control over the electronic and magnetic interactions 
is to produce magnetic semiconductor superlattices. These are the digital ferromagnetic 
heterostructures (DFH), which are obtained by $\delta$-doping with Mn a low 
temperature GaAs MBE-grown layer \cite{Kaw1}. 
Here Mn concentrations as high as 50\% are obtained in a few (typically 2-3) GaAs 
monolayers and one may expect correspondingly higher large Curie
temperatures. However these structures show properties rather different from those
of their random alloy counterparts. Here we report the most relevant experimental
findings.

\vspace{0.1in}

1) The Curie temperature is rather low ($\sim50$~K) and independent of the 
separation between the Mn planes \cite{Kaw1}. 
$T_c$ usually decays with increasing GaAs layer thickness 
separating the MnAs sub-monolayers, and saturates for thicknesses larger than 
$\sim$50 GaAs monolayers. The saturation is unexpected according to the mean 
field model for three dimensional systems, since the total Mn concentration in 
the sample decreases with the increase of the GaAs thickness \cite{Dietl}. 
This separation dependence suggests that DFH behave like planar systems.

\vspace{0.1in}

2) Hall measurements in the direction parallel to the MnAs planes show 
an anomalous Hall effect for undoped samples, which disappears upon Be doping 
\cite{gwin1,gwin2}. Large Shubnikov de Haas oscillations are found in 
doped samples, although surprisingly the carrier densities
extracted from the Hall coefficient and from the Shubnikov de Haas oscillations
are different. This suggests that two different carrier types could
be present in the system. 

\vspace{0.1in}

3) There is a correlation between metallicity and Curie temperature. In low $T_C$
samples the transport is given by activated hole conduction and this is consistent 
with variable range hopping in two dimensions \cite{Luo,mc2,Ted}. In contrast, 
in the only DFH to date showing $T_C$ around room temperature 
the transport is $p$-type and metallic \cite{McCombe}. However 
in these latter structures made from (Ga,Mn)Sb two phases may be present \cite{mc2,McCombe}, 
with a diluted phase responsible for the ferromagnetism below 40~K and a zincblende 
MnSb phase responsible for the room temperature ferromagnetism.

\vspace{0.1in}

4) GaAs/AlAs band engineering and spatial selective doping \cite{Tanaka,Furdyna}
allow the enhancement of the $T_C$ in AlAs/(Ga,Mn)As DFH with respect to their
GaAs/(Ga,Mn)As counterparts. This enhancement is correlated with the enhanced hole
concentration in the Mn layers and to the achievement of metallic conductance.

\vspace{0.1in}

Finally we point out that some of the aspects described above are common
to other $\delta$-doped structures. For instance it has been recently demonstrated
\cite{Noh} that Be $\delta$-doped low temperature GaAs undergoes an insulator to
metal transition as the Be concentration is enhanced. In this case the transport
changes from $n$-type thermally activated, to $p$-type metallic. The first is
reported for small Be concentrations and is due to the hopping between As antisite
levels in the GaAs region, while the second dominates at large Be concentrations and 
is due to extended hole states in the Be rich region.

From this brief overview it is clear that DFH present rather rich and complex
physics, which calls for an extensive theoretical analysis. So far the magnetic
properties of DFH have been studied only within the mean field approach 
\cite{Meyer,Sham}, while the transport has been investigated solely in the 
ballistic limit for the case of 100\% Mn doping in the plane \cite{usprl}. 
Here we report an
extensive {\it ab-initio} study of the electronic, magnetic and transport
properties of both GaAs/(Ga,Mn)As and AlAs/(Ga,Mn)As DFH, for various Mn
concentrations and As antisite doping levels. 

The paper is organized as follows. In the next section we describe our
computational technique and we motivate the approximations made. Then
we investigate the properties of GaAs/(Ga,Mn)As DFH, the effects of As
antisites, and the properties of AlAs/(Ga,Mn)As DFH. Finally we conclude and we
suggest new ways to manipulate the properties of DFH.

\section{Computational Technique}

We perform density functional theory (DFT) \cite{kohn} calculations within the 
local spin-density approximation (LSDA). The use of LSDA for DMS is very well 
documented \cite{usreview}, and it provides a good description of
the main physics of (Ga,Mn)As. Recently we have demonstrated \cite{alessio} that
self-interaction corrections to the LSDA do not strongly affect the band structure of
(Ga,Mn)As, although they lead to strong localization and orbital ordering of the
Mn $d$ shell in (Ga,Mn)N. For this reason we choose to work within the LSDA.

Our numerical implementation, contained in the code SIESTA 
\cite{Siesta1,Siesta2}, uses pseudopotentials and a highly
optimized localized atomic orbital basis set. These two aspects make
SIESTA extremely suitable for handling systems with a large number of atoms in the
unit cell without a significant loss of accuracy. The drawback is that both the 
pseudopotentials and the basis set must be accurately optimized. 

First we consider the pseudopotentials. We use well-tested scalar 
relativistic Troullier-Martins pseudopotentials \cite{TM} with non linear 
core corrections \cite{Lou82} and Kleinman-Bylander factorization \cite{KB1}. 
The eigenvalues for the valence electrons of the free atom are compared with
those generated for an all-electron calculation for different atomic and
ionic configurations. Then we perform total energy calculations for elementary
solids comparing the lattice constant, the bulk modulus and the band structure
with reference calculations. These are performed with a well-converged basis
set. Note that this is quite a delicate procedure, since with localized orbital
basis sets the variational principle is not governed by a single-parameter 
such as the cut-off energy with plane-waves.

Finally an optimized basis set is selected. The basis functions in
SIESTA are the product of an angular function with a given
angular momentum, and a radial numerical function. This latter is constructed as the 
DFT solution of the free pseudo-atom with an additional hard-wall confining potential.
Furthermore, in order to enhance the variational freedom, several radial functions 
(``zetas'') for the same angular momentum are constructed with the `split valence' scheme 
\cite{emilio}. In the case of (Ga,Mn)As the crucial aspect is
to introduce several zetas for the Mn $d$ orbitals, and the criterion
we have adopted is that of reproducing the physics of the $d$ shell of Mn in
MnAs. More details are given in reference \cite{usprb}. Here we only mention
that the same procedure has been adopted for the Al pseudopotential and basis
set. These have been checked for both metallic Al and zincblende AlAs. The reference
configuration for the pseudopotential is 3s$^2$3p$^1$ with cut off radii 1.90, 
and 1.80 a.u. respectively for the $s$ and $p$ shells. 
Finally the basis set for Al has two basis functions 
for both $s$ and $p$ electrons, with the same cut-off radius of 6.0 a.u. and 
the `split norm' parameter is 0.15 \cite{Siesta2}.

\begin{figure}[ht]
\epsfxsize=4cm
\centerline{\epsffile{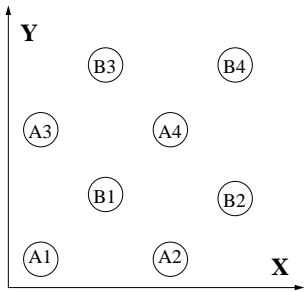}}
%\epsfxsize=6cm
%\centerline{\epsffile{TEST}}
\caption{The eight possible positions of the Mn ions in the supercell.} 
\label{F1}
\end{figure}

All the calculations presented here are performed within a supercell scheme. Our
supercell is constructed from a $2\times2\times3$ zincblende cubic cell (lattice constant
$a_0$=5.65\AA) and contains
96 atoms in total. We mimic a DFH by replacing Ga with Mn ions only in one of the 
GaAs planes. We use periodic boundary conditions in all directions sampling 
18 $k$-points in the supercell Brillouin zone. This corresponds to a 
(Ga,Mn)As/GaAs superlattice in which the (Ga,Mn)As planes are separated by 6 GaAs monolayers
(16.95\AA). In this supercell the Mn ions can occupy only eight possible positions 
in plane. These are arranged into two simple cubic lattices
translated with respect to each other along the diagonal of the $xy$ plane. 
Since the exact positions of the Mn ions is rather important in determining the electronic
structure, these are schematically presented in figure \ref{F1}.

\section{(G\lowercase{a},M\lowercase{n})A\lowercase{s} DFH}

In this section we analyze the effects of the Mn concentration on the electronic
properties of (Ga,Mn)As DFH, by calculating the band structure, the DFT total potential,
the strength of the FM coupling and the transport properties. Our main
aim is to monitor the evolution of these quantities as a function of the Mn 
concentration in the (Ga,Mn)As plane. In particular we want to establish
whether there is a correlation 
between the Mn concentration and the metallicity of the system.

\subsection{Band structure}

\begin{figure}[ht]
\epsfxsize=8cm
\centerline{\epsffile{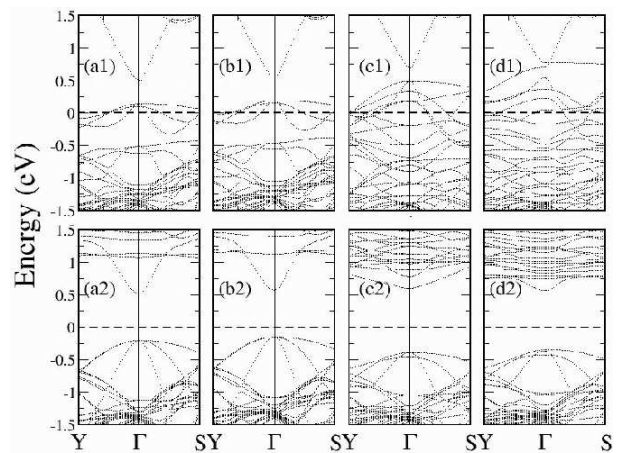}}
%\epsfxsize=6cm
%\centerline{\epsffile{TEST}}
\caption{Band structure of the (Ga,Mn)As/GaAs DFH described in the text as a function of the
Mn concentration. The directions are parallel to the MnAs planes.
The upper panels correspond to the majority spin-band and the lower to 
the minority. The Mn concentrations are a) 12.5\%, b) 25\%, c) and d) 50\%. In the case of
50\% concentration we consider two different arrangement of the Mn atoms: c) A1, A2, A3, A4,
d) A1, A2, A3, B1.} 
\label{F2}
\end{figure}

In figure \ref{F2} we present the band structure of our (Ga,Mn)As/GaAs DFH 
for Mn concentrations of 12.5\%, 25\% and 50\%. In figures 
\ref{F2}a, \ref{F2}b and \ref{F2}c, only the sites belonging to one 
of the two cubic sub-lattices in the plane are occupied. This maximizes the 
mean Mn-Mn separation. In contrast in figure \ref{F2}d sites 
belonging to both the lattices are occupied (namely A1, A2, A3, and B1).
Here we
plot the bands only along two directions parallel to the (Ga,Mn)As plane,
since in the perpendicular direction these are very similar to the case of
MnAs planes embedded in GaAs \cite{usprl} and they do not change
significantly with the Mn concentration.
In contrast for directions parallel to the Mn planes, 
we expect a transition from the band structure of a (Ga,Mn)As random alloy
to the metallic band structure of a zincblende MnAs plane \cite{usprl}. This is indeed 
the behavior observed in figure \ref{F2}. If one increases the Mn concentration 
from 12.5\% to 50\% (from (a) to (b) to (c)) the Fermi level shifts 
downward in energy ending up deep in the majority spin valence band. 
At the same time also the spin-splitting of the valence band increases. Note that 
the DFH behaves as an half-metal at every concentration. 
The magnetic moment of the supercell therefore is always $4\mu_\mathrm{B}\times N_\mathrm{Mn}$ 
where $N_\mathrm{Mn}$ is the number of Mn atoms in the cell. M\"ulliken
population analysis \cite{usprb,Mul} shows an orbital population for the Mn $d$ shell 
of $\sim4.7$ for the majority spin electrons and of $\sim0.8$ for the minority, although both
depend on the specific spatial arrangement of the Mn ions. These aspects are
consistent with the picture of Mn as a single acceptor in GaAs. The Mn ions are in a
$d^5$ state with an associated antiferromagnetically coupled hole, as for the 
case of the random alloys \cite{usprb}.

However
the situation of figure \ref{F2}d is in stark contrast with this picture. 
In this case the Mn concentration is still 50\% but three of the four Mn 
ions in the plane occupy nearest neighbor positions. The band gap in the 
majority spin band closes and the material is an half metal with a completely 
metallic majority spin band. This suggests that the actual position of the Mn 
atoms in the plane is crucial in determining the electronic properties. The same 
sensitivity of the electronic structure to the position of the Mn ions is also 
present in the random alloys \cite{usapl}. However this generally does not lead
to a strong distortion of the band structure, and the bands obtained for Mn
ions diluted in the supercell or occupying nearest neighbor positions are quite
similar. In DFH the planar arrangement of the Mn ions makes the system 
more confined, and therefore more sensitive to inhomogeneities. This of course
drastically affects the scattering properties of electrons in the (Ga,Mn)As
planes, as we will show in the following sections.

{\subsection{DFT potential}}

One fundamental question for understanding the physics of DFH is: ``are the 
spin carriers confined in the (Ga,Mn)As plane or do they spread over the 
GaAs spacer?''. In order to
answer this question it is useful to investigate the behavior of the total 
DFT potential along the superlattice direction $z$. This of course has the same
periodicity as the atomic lattice. However we are not interested in the
potential at the atomic scale, and instead we perform a ``macroscopic 
average'' \cite{Bald}. The macroscopic average is obtained by first taking a
planar average and then by averaging the result over the period of a GaAs
monolayer along the superlattice direction. 
The resulting $z$-dependent potential is that felt by an electron 
with long wave-length at a density small enough not to perturb significantly 
the potential. Therefore this can be interpreted as the effective 
$z$-dependent potential in the spirit of the envelope function approximation.

\begin{figure}[ht]
\epsfxsize=9cm
\centerline{\epsffile{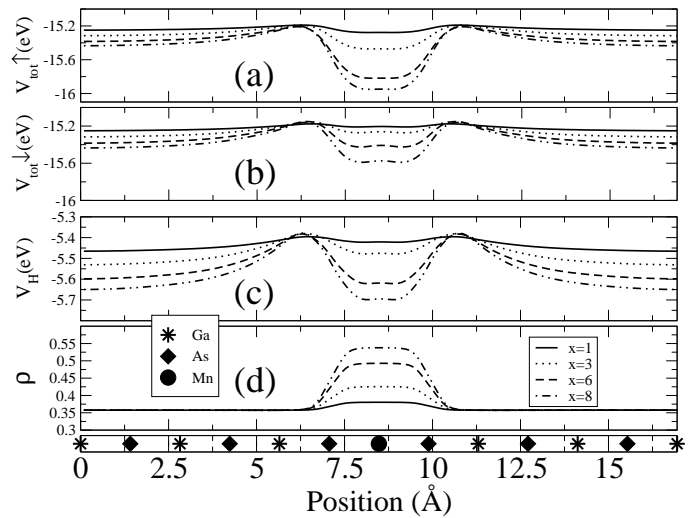}}
%\epsfxsize=6cm
%\centerline{\epsffile{TEST}}
\caption{Macroscopic average of a) the total DFT potential for majority spins,
b) the total DFT potential for minority spins, c) the Hartree potential and the d) charge
density distribution as a function of the position along the superlattice
direction, for different Mn concentrations ($x$ is the number of Mn ions in the plane).} 
\label{F3}
\end{figure}

In figure \ref{F3} we plot the macroscopic average for the total DFT potential, 
the Hartree potential and the charge
density distribution as a function of the position along the superlattice
direction for different Mn concentrations.
Notice first that there is indeed a confining potential in the 
Mn plane. The charge density is unevenly distributed along the
superlattice and accumulates in the Mn plane. This is consistent
with the picture of the Mn ions that we have given in the past \cite{usprb}, 
in which the majority spin hole is nearly bound to the Mn ion while
the minority feels a much weaker potential.

\begin{figure}[ht]
\epsfxsize=4.5cm
\centerline{\epsffile{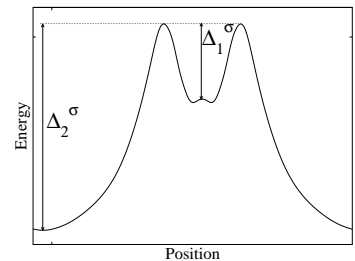}}
%\epsfxsize=6cm
%\centerline{\epsffile{TEST}}
\caption{Definition of $\Delta_1^\sigma$ and $\Delta_2^\sigma$.} 
\label{F4}
\end{figure}

The $z$-dependent total potential is strongly 
spin-dependent. Generally it has a double well structure, with two potential 
minima, located in the (Ga,Mn)As plane and in the GaAs spacer respectively. 
These are separated by a potential barrier which grows if the Mn concentration 
in the plane increases. We further investigate the nature of the confining
potential by calculating its evolution upon increasing the Mn doping. For this
purpose we define $\Delta_1^\sigma$ and $\Delta_2^\sigma$ as the energy 
minima in in the (Ga,Mn)As plane and the GaAs spacer respectively 
measured with respect to the top of the energy barrier (see figure \ref{F4}).
$\sigma$ is the spin index and $\uparrow$ ($\downarrow$) indicates the 
majority (minority) spin electrons.
We denote  as $\Delta_1^\mathrm{H}$ and $\Delta_2^\mathrm{H}$
the same quantities for the Hartree potential.

\begin{figure}[ht]
\epsfxsize=8cm
\centerline{\epsffile{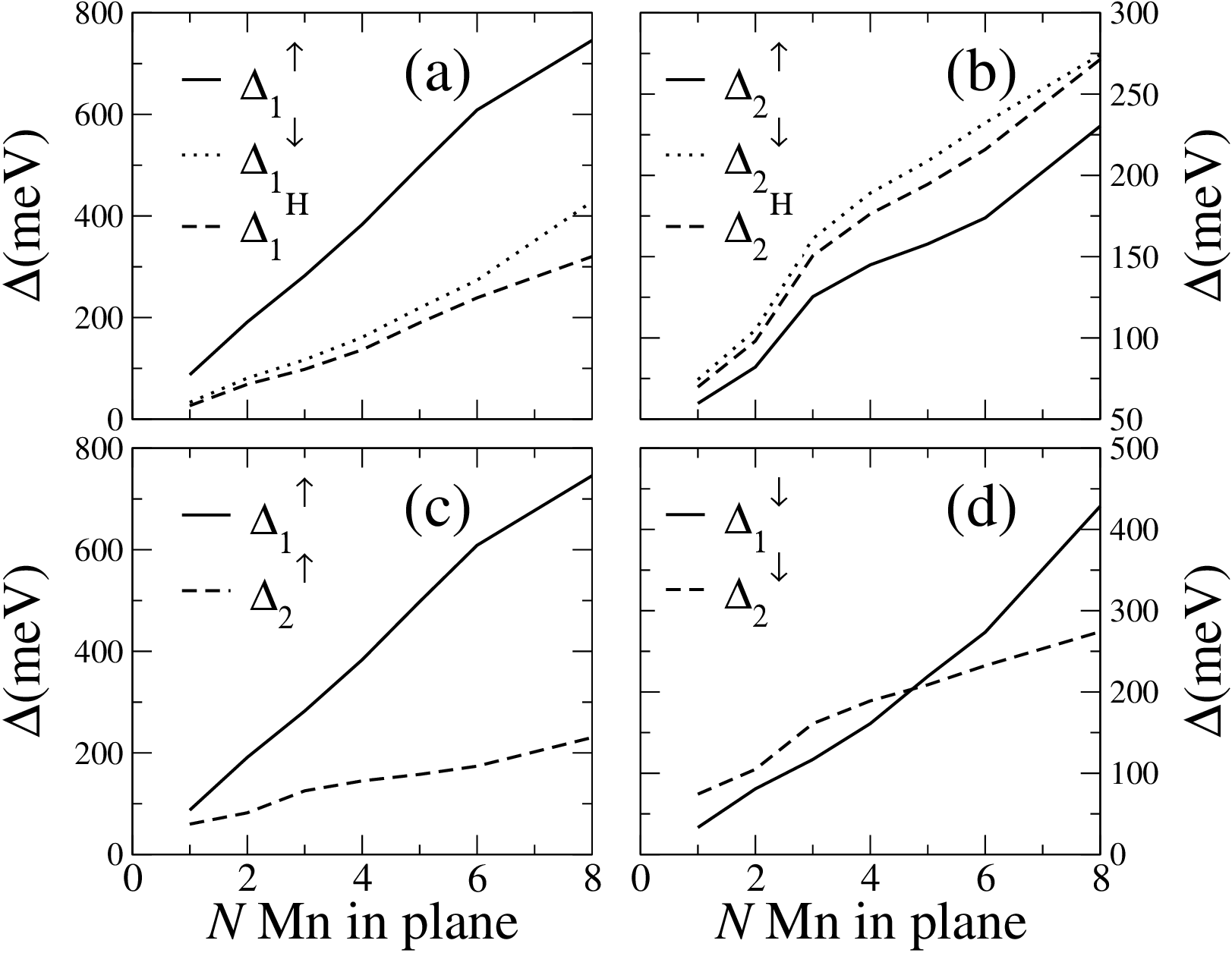}}
%\epsfxsize=6cm
%\centerline{\epsffile{TEST}}
\caption{$\Delta_1^\sigma$, $\Delta_2^\sigma$,
$\Delta_1^\mathrm{H}$ and $\Delta_2^\mathrm{H}$ as a function of the Mn
concentration in the (Ga,Mn)As plane.} 
\label{F5}
\end{figure}

In figure \ref{F5} we present these quantities as a function of the number of Mn
ions in the Mn plane (the maximum number is 8). First we
notice that for minority spins both $\Delta_1^\downarrow$ and
$\Delta_2^\downarrow$ mimic closely the behavior of the Hartree 
potential (figure \ref{F5}a, and \ref{F5}b).
This means that the DFT potential for the minority spins is largely electrostatic 
with small contributions from the exchange part. In contrast 
$\Delta_1^\uparrow$ for the majority electrons deviates strongly from the
behavior of its electrostatic component, and the deviation increases as the Mn
concentration is enhanced. 

If we now compare $\Delta_1$ with $\Delta_2$ for the two spin species we find a
rather different behavior. For the majority spin electrons (figure \ref{F5}c) 
$\Delta_1>\Delta_2$ at all Mn concentrations. Furthermore the difference 
$\Delta_1-\Delta_2$ increases with increasing the Mn concentration. This means
that the minimum in the (Ga,Mn)As plane is always the lower of the minima for majority
electrons and it becomes more stable as the Mn concentration increases.

In contrast, for the minority electrons there is not a striking difference
between $\Delta_1$ and $\Delta_2$ meaning that they are less confined in the
(Ga,Mn)As plane (see figure \ref{F5}d). It is also interesting to point out that 
$\Delta_2>\Delta_1$ at low Mn concentration but there is a cross over 
for doping of the order of 50\%. This is suggestive of the fact that the
confinement of minority electrons switches from the GaAs to the (Ga,Mn)As
region upon the increase of Mn concentration. Of course there are no empty minority
spin states at the Fermi level and therefore in these structures the minority spins
contribute little to the electron transport. However in the case of additional
doping (for instance by intrinsic As antisites, As$_\mathrm{Ga}$) an impurity band 
may form at
the Fermi level, opening a transport channel in the minority sub-band. The
potential described here suggests that those electrons will hardly distribute
over
the (Ga,Mn)As plane at low doping but they will invade the (Ga,Mn)As region at
large Mn doping. For this reason we expect a spatial separation of the spin
currents (majority spin in the (Ga,Mn)As plane, and minority spin in the GaAs
spacer) for low concentrations, and strong spin-mixing at higher concentrations.

\subsection{Stability of the ferromagnetic state}

In this section we investigate the strength of the ferromagnetic
coupling in (Ga,Mn)As DFH, and in particular we discuss how the coupling
depends on Mn doping. The relevant quantity to investigate is
the energy difference $\Delta_\mathrm{FA}=E_\mathrm{AF}-E_\mathrm{FM}$ between 
the total energy of the antiferromagnetic ($E_\mathrm{AF}$) and the 
ferromagnetic ($E_\mathrm{FM}$) configurations of the supercell. 
Here we investigate the case of 25\%, 37.5\% and 50\% Mn doping in the plane, 
where respectively two, three and four Mn ions are present in the
supercell. This gives us some freedom to arrange the Mn ions in different 
ways and to investigate different antiferromagnetic configurations. A summary
of the cases studied is presented in table \ref{T1}.
\begin{table} [ht]
\caption{Summary table of the different magnetic configurations studied. 
The labels for the positions of the Mn ions 
are those of figure \ref{F1}. \label{T1}}
\centering\begin{tabular}{ccccccc} \hline\hline
\multicolumn{2}{c} [Mn]  Configuration & \multicolumn{5}{c} {Position}  \\ 
  & & A1 & A2 &  A3 & A4 & B1 \\ \hline
25\% & FM$_1$   & $\uparrow$ & - & - & - & $\uparrow$ \\   
25\% & AFM$_1$   & $\uparrow$ & - & - & - & $\downarrow$ \\   
25\% & FM$_2$   & $\uparrow$ & $\uparrow$ & - & - & - \\   
25\% & AFM$_2$   & $\uparrow$ & $\downarrow$ & - & - & - \\   
25\% & FM$_3$   & $\uparrow$ & - & - & $\uparrow$ & - \\   
25\% & AFM$_3$   & $\uparrow$ & - & - & $\downarrow$ & - \\ \hline
37.5\% & FM$_1$   & $\uparrow$ & - & $\uparrow$ & $\uparrow$ & - \\   
37.5\% & AFM$_1$   & $\uparrow$ & - & $\uparrow$ & $\downarrow$ & - \\   
37.5\% & AFM$_2$   & $\uparrow$ & - & $\downarrow$ & $\uparrow$ & - \\   
37.5\% & FM$_{1n}$   & $\uparrow$ & - & $\uparrow$ & - & $\uparrow$  \\   
37.5\% & AFM$_{1n}$   & $\uparrow$ & - & $\uparrow$ & - & $\downarrow$  \\ \hline 
50\% & FM$_1$   & $\uparrow$ & $\uparrow$   & $\uparrow$   & $\uparrow$ & - \\ 
50\% & AFM$_1$  & $\uparrow$ & $\downarrow$ & $\downarrow$ & $\uparrow$ & - \\ 
50\% & AFM$_2$  & $\uparrow$ & $\downarrow$ & $\uparrow$   & $\uparrow$ & - \\ 
50\% & AFM$_3$  & $\uparrow$ & $\uparrow$   & $\downarrow$ & $\downarrow$ & - \\ 
50\% & FM$_{1n}$  & $\uparrow$ & $\uparrow$  & $\uparrow$ & - & $\uparrow$   \\ 
50\% & AFM$_{1n}$ & $\uparrow$ & $\uparrow$  & $\uparrow$ & - & $\downarrow$   \\ 
50\% & AFM$_{2n}$ & $\downarrow$ & $\uparrow$ & $\uparrow$ & - & $\uparrow$   \\ 
50\% & AFM$_{3n}$ & $\uparrow$ & $\uparrow$ & $\downarrow$ & - & $\downarrow$   \\ \hline\hline
\end{tabular}
\end{table}

We ran a total energy calculation for each of the configurations of table
\ref{T1}, and fit our calculations to a third nearest neighbors Heisenberg 
model, in which the total energy $E$ is expressed as
\begin{equation}
E=-\sum_{i<j}^{\mathrm{nn}}J_{ij}\vec{S}_i\cdot\vec{S}_j\;.
\end{equation}
$J_{ij}$ are the exchange constants, $\vec{S}_i$ is the spin of the
$i$-th Mn ions, and the sum runs up to third nearest neighbors. 
The results of this fit are presented in table \ref{T2}, where by convention we
use $|\vec{S}|=5/2$. Note that in our fitting procedure we have more configurations
than parameters, and it is remarkable to observe that the fit is generally very
good.
\begin{table} [ht]
\caption{$J_1$, $J_2$ and $J_3$ from the total energy calculations for the
configurations of table \ref{T1}. The units are meV. \label{T2}}
\centering\begin{tabular}{c|cccc} \hline\hline
[Mn] in plane & $J_1$ & $J_2$ & $J_3$ &  \hspace{0.1in} $x\sum_iJ_{i}$ \\ \hline
$x=0.25$  & 23.3 & 2.9 & 5.6 & 7.95 \\
$x=0.375$ & 19.8 & 1.4 & 4.9 & 9.8 \\
$x=0.50$  & 13.3 & 0.9 & 4.5 & 9.35 \\ \hline\hline
\end{tabular}
\end{table}
From the table it is clear that the first nearest neighbor coupling constant,
$J_1$, is responsible for most of the coupling which decays rapidly with the Mn-Mn
separation. It is also interesting to note that the second nearest neighbor
coupling, $J_2$, is small for all the concentrations studied.
Remarkably, all the coupling constants are strongly dependent on the Mn 
concentration in the plane, and they decay quickly as this is increased.
This decay, which is particularly severe for the first nearest neighbor
coupling constant, is in stark contrast with the expectations of RKKY-like theories
in which the $J$'s increase as $x^{1/3}$ at zero temperature \cite{MCD1}.
However it is consistent with previous DFT-LDA calculations \cite{Mark}
for DMS random alloys, and illustrates once again the critical dependence of the
Mn-Mn ferromagnetic coupling on the local chemical environment.

In order to put our calculations in perspective, next we estimate the 
ferromagnetic Curie temperature, $T_C$, of the system for different Mn
concentrations. This can be simply obtained by using the mean field expression 
for a three dimensional Heisenberg model, which reads
\begin{equation}
T_Ck_B=\frac{2}{3}S^2n_\mathrm{Mn}\sum_iJ_{i}\;, 
\label{tcmf}
\end{equation}
where $n_\mathrm{Mn}$ is the Mn concentration, and the sum runs over all the
cation sites. This is a rather crude approximation and a complete 
thermodynamic theory should include the elementary 
spin-excitations \cite{MacDonspex,Bhattspex}. Moreover in the present case the 
Mn concentration is not a well-defined quantity since the (Ga,Mn)As region 
cannot be separated from the GaAs region. Roughly speaking one should consider 
the volume of (Ga,Mn)As to be the region around the Mn ions as thick as the 
range of the relative 
confining potential (see figure \ref{F3}). This quantity is not clearly 
defined. Therefore, assuming that the (Ga,Mn)As volume does not depend
on the Mn concentration, we prefer to evaluate only the following ``magnetic
energy''
\begin{equation}
E_\mathrm{mag}=x\sum_iJ_{i}\;, 
\label{magen}
\end{equation}
which is proportional to the Curie temperature.

From table \ref{T2} one notes that $E_\mathrm{mag}$ has a non-monotonic
dependence on the Mn concentration, presenting a maximum for $x=0.375$. 
Such behavior is generally observed in DMS random alloys, for which there is a maximum 
of $T_C$ upon Mn doping, followed by a sharp decay for large Mn concentrations 
(above $x=0.05$) \cite{Ohno2,Ohno3}. This usually coincides with the loss of the metallic
state. In contrast, in DFH made to date the 
situation seems to be reversed \cite{Kaw1}, with a larger $T_C$ for larger Mn 
concentrations. This apparently contradicts our predictions.
However two important aspects need to be considered. First in actual DFH the
transport is through variable range hopping \cite{Ted}, while in our supercell
calculations the system is ``by definition'' metallic. Secondly 
DFH usually present very strong compensation. This indicates that 
a large number of donors, whose density is probably related to the in plane Mn
concentration, are present in the DFH. In the next sections we will investigate
systematically both the transport properties and the effects of the presence of
donors.

\subsection{Ballistic Transport}

In this section we investigate the ballistic transport in DFH, with the aim
of understanding the nature of the electronic states responsible for the conductance.
The technique used is identical to that described in references
\cite{usprl,uslb} and here we summarize only the main aspects. The transport is 
calculated by using the self consistent tight-binding like Hamiltonian
and overlap matrix computed by SIESTA. The matrix elements are obtained
from the self-consistent charge density by evaluating numerically both two-
and three-center integrals \cite{Siesta2}. Then we rewrite both the Hamiltonian
and the overlap matrix in a tridiagonal form along the direction of the transport and
we use periodic boundary conditions along the other directions. 
Finally, the $k$-dependent transmission matrix $t_\sigma(k)$ for the spin
direction $\sigma$ is calculated by using our Green's function technique 
\cite{uslb}, and the spin conductance $\Gamma^\sigma$ in the 
Landauer-B\"uttiker formalism \cite{lb},
\begin{equation}
\Gamma^\sigma=\frac{e^2}{h}\sum^\mathrm{BZ}_k\mathrm{Tr}\;t_\sigma(k)t_\sigma(k)^\dagger\;,
\label{lb}
\end{equation}
where we integrate over the two-dimensional Brillouin zone in the 
plane orthogonal to the direction of the transport.
Here we consider a two spin fluid model, where there
is no mixing between the majority and minority spin currents.

\begin{figure}[ht]
\epsfxsize=7cm
\centerline{\epsffile{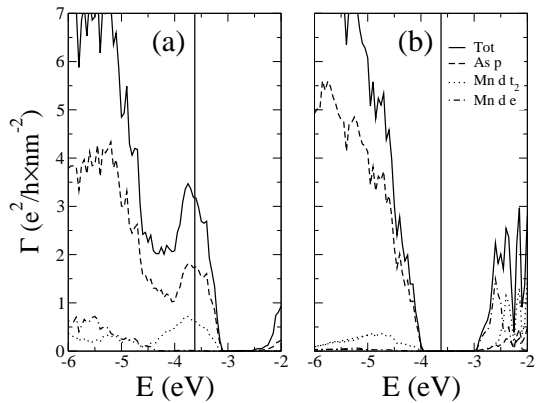}}
%\epsfxsize=6cm
%\centerline{\epsffile{TEST}}
\caption{Conductance as a function of energy for a 50\% Mn DFH 
in the FM state. The Mn ions are uniformly spaced in the plane
(configuration FM$_1$ of table \ref{T1}). The vertical line indicates
the position of the Fermi level. (a) majority, (b) minority spins.} 
\label{F6}
\end{figure}

We study transport only in the direction parallel to the Mn plane. In the
orthogonal direction in fact the transport is mainly due to hopping between the 
Mn planes and it is strongly suppressed if these are sufficiently far apart 
\cite{usprl}. In figure \ref{F6} we present the conductance as 
a function of energy for a 50\% Mn supercell respectively in the FM
state, with the Mn ions uniformly distributed in the plane (configuration
FM$_1$ of table \ref{T1}).

As expected from the band structure (figure \ref{F2}c) the conductance as a function of
energy shows an half-metallic behavior, with zero conductance for the minority spin
band at the Fermi level. This 100\% spin polarization of the conductance persists down to 0.5~eV
from the Fermi energy, where minority states at the top of the valence band start contributing
to the current. Turning our attention to the orbital contribution to the conductance it is important
to observe that, at the Fermi level the current is entirely due to a mixture of As $p$ and Mn $d$
$t_2$ states. It is also interesting to note that non-negligible Mn $d$ contributions are
present in the majority spin band for energies down to 4~eV below the Fermi level. This is an
indication of the strong $p$-$d$ hybridization in the majority spin band, and in particular at
its top. 
In contrast the Mn $d$ contribution to the conductance is almost negligible in the valence
minority band, and Mn $d$ states appear only for $E\ge-3$~eV, relative to the bottom of
the conduction band.

This 100\% spin-polarization at the Fermi level is very encouraging for the potential
use of DFH as spin-injector
for spintronics devices. However it is crucial to investigate how this feature survives when
compensating defects are present in the system.

\section{(G\lowercase{a},M\lowercase{n})A\lowercase{s} DFH:
effects of A\lowercase{s}$_\mathrm{G\lowercase{a}}$}

As in the case with the (Ga,Mn)As random alloys, DFH are also usually strongly
compensated so that the hole concentration is considerably lower than the Mn concentration.
It is generally accepted that the strong compensation is due to donors, most
likely of intrinsic defects. In particular both As antisites (As$_\mathrm{Ga}$) 
\cite{Ohno3,usapl} and interstitial Mn (Mn$_i$) \cite{mninter1,mninter2}
have been indicated as relevant compensating defects. The relative abundance of those donors
probably depends on the growth conditions, the Mn concentration, and the post-growth processing.
Since DFH are usually grown under large As overpressure \cite{Kaw1} we believe that in this
case As antisites dominate. Here we investigate how the electronic properties of a 50\% Mn DFH
changes upon As$_\mathrm{Ga}$ doping.

\subsection{Band Structure}

\begin{figure}[ht]
\epsfxsize=8cm
\centerline{\epsffile{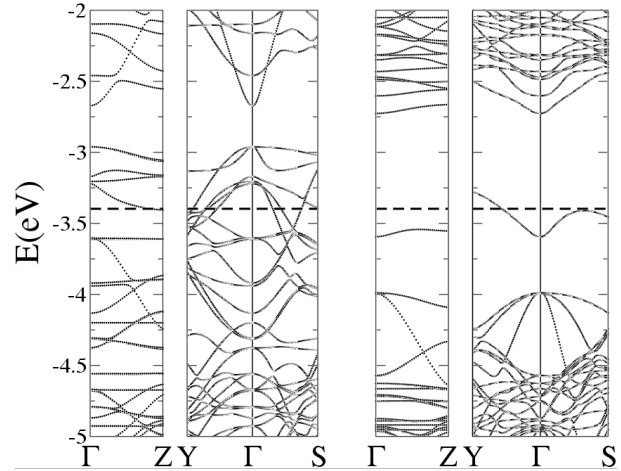}}
%\epsfxsize=6cm
%\centerline{\epsffile{TEST}}
\caption{Band structure for 50\% Mn DFH where a single As$_\mathrm{Ga}$ is 
introduced in the GaAs spacer at midway between two (Ga,Mn)As planes. The
directions are: 1) Y-$\Gamma$-S in the (Ga,Mn)As plane, 2) $\Gamma$-Z perpendicular to the
(Ga,Mn)As plane. The horizontal line indicates the position of the Fermi level.
On the left (right) we plot the majority (minority) band.} 
\label{F7}
\end{figure}

As in the previous section the band structure provides important information on the 
electronic properties of the DFH. In figure \ref{F7} we present the band
structure for 50\% in plane Mn DFH, where a single As$_\mathrm{Ga}$ is introduced
into the GaAs spacer at midway between two consecutive (Ga,Mn)As planes
(the total As$_\mathrm{Ga}$ concentration is $\sim$2\%). 

The most important feature of this band structure is that the half-metallic
state is destroyed. This is due to the appearance of the As$_\mathrm{Ga}$
level in the minority spin band, and to the fact that the Fermi energy is pined at
this level. An As antisite in GaAs is a double donor with a doubly occupied 
deep level at midgap and an empty resonant state at the edge of the conduction
band. These states have respectively the A and T$_2$ symmetries of the 
$T_\mathrm{d}$ point group \cite{usjmmm}. In (Ga,Mn)As the density of such
defects is generally rather large and they can give rise to the formation of 
narrow impurity bands. In addition in the case of DFH it is likely that the As 
antisites will concentrate in the proximity of the Mn layers. This is confirmed
by our total energy calculations, which show that there is an energy gain of
approximately 0.9~eV when an As antisite moves from the middle of the cell to the 
Mn plane.

The total magnetic moment of such a unit cell 
is about 17.2$\mu_\mathrm{B}$ and depends weakly on the position
of the As antisite with respect to the Mn plane (it is 17.33$\mu_\mathrm{B}$ when
the As$_\mathrm{Ga}$ lies in the (Ga,Mn)As plane). 
Assuming a rigid band model and considering that, in absence of 
As antisites the magnetic moment of the cell is 16$\mu_\mathrm{B}$, 
we conclude that an As$_\mathrm{Ga}$ contributes 0.4 electrons to the minority band 
and 1.6 to the majority (we recall that one As$_\mathrm{Ga}$ also introduces an impurity 
state that can accommodate two electrons).
Therefore the presence of 
one As antisite in such a unit cell has two main effects: i) it compensates up to 1.6 
holes in the majority spin band, and ii) it opens a conduction channel at the Fermi
level in the minority spin band. This is a crucial aspect for understanding the 
transport properties of such structures.

\subsection{Transport in presence of As antisites}

\begin{figure}[ht]
\epsfxsize=7cm
\centerline{\epsffile{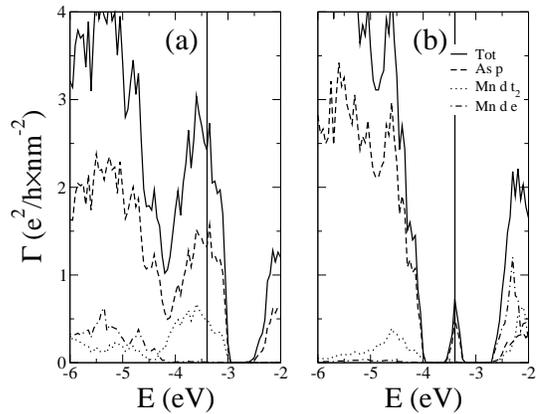}}
%\epsfxsize=6cm
%\centerline{\epsffile{TEST}}
\caption{Ballistic conductance as a function of energy for a 50\% Mn DFH with one
As$_\mathrm{Ga}$ in the middle of the unit cell, half the way between two
consecutive (Ga,Mn)As planes. The vertical line indicates the position of the
Fermi level. (a) majority, (b) minority spins.} 
\label{F8}
\end{figure}

We calculate the in plane ballistic conductance for a 50\% Mn DFH 
with one As$_\mathrm{Ga}$ in the middle of the spacer and the Mn ions 
uniformly distributed in the plane (configuration FM$_1$). 
This is the same situation as in figure \ref{F6}. The results are presented in
figure \ref{F8}, where again we have considered up to 100 $k$-points
in the transverse Brillouin zone.

The main difference with respect to the As$_\mathrm{Ga}$-free case is the 
presence of a non-vanishing conductance for the minority spin electrons at the
Fermi energy. This is entirely due to the As$_\mathrm{Ga}$ impurity band as pointed
out in the previous section, and of course destroys the half-metallic behavior
of the spin current. The spin-polarization of
the current $P$, defined as 
\begin{equation}
P=\frac{\Gamma^\uparrow-\Gamma^\downarrow}
{\Gamma^\uparrow+\Gamma^\downarrow}\;,
\end{equation}
where the conductances are taken at the Fermi level, drops from 100\% in the 
defect-free case to about 60\% in the present case.

However the situation is rather different if we consider diffusive transport 
\cite{mazin}. This is due to the different way in which the density of states 
and the group velocity contribute to the conductance in the ballistic and 
diffusive limits.
In fact, in the ballistic limit the conductance of a uniform
system can be obtained simply by summing up the number of scattering channels 
at the Fermi level
\begin{equation}
\Gamma^\sigma=\frac{e^2}{h}\sum_k^\mathrm{BZ}1=
\frac{e^2}{h}\sum_k^\mathrm{BZ}N_k^\sigma v_k^\sigma=
\frac{e^2}{h}\langle N {v}\rangle^{\sigma}\;,
\end{equation}
where $N_k^\sigma$ and $v_k^\sigma$ are respectively the density of state and the group
velocity for a spin $\sigma$ scattering channel, and the sum is performed over 
the two dimensional Brillouin zone orthogonal to the transport direction.
In contrast from the classical Boltzmann equation one finds that the 
diffusive spin conductance is proportional to
\begin{equation}
\tau^\sigma\langle N {v}^2\rangle^{\sigma}=\tau^\sigma\sum_k
N_k^\sigma {v_k^\sigma}^2
\;,
\end{equation}
where now the sum runs over the three dimensional Fermi surface and 
$\tau^\sigma$ is the spin-dependent relaxation time \cite{mazin}.
This reflects the well-known fact that, while in the ballistic limit
all the scattering channels contribute with $e^2/h$ to the conductance 
independently from their group velocity, in the diffusive case the current
is dominated by fast electrons because of the $v^2$ dependence.

If one assumes that the relaxation time is not dependent on the
spin direction ($\tau^\uparrow=\tau^\downarrow$), then
the spin polarization of the current in the diffusive limit can
be written as
\begin{equation}
P_{\langle N {v}^2\rangle}=\frac{\langle N {v}^2\rangle^{\uparrow}-\langle N {v}^2\rangle^{\downarrow}}
{\langle N {v}^2\rangle^{\uparrow}+\langle N {v}^2\rangle^{\downarrow}}\;.
\end{equation}
In general $P$ and $P_{\langle N {v}^2\rangle}$ are different, with 
$P_{\langle N {v}^2\rangle}$ larger if the difference in conductance for the two
spin channels originates from a large Fermi velocity mismatch between the two
spin bands. This is the case in the present DFH. In the majority band the Fermi
surface is derived from the top of the GaAs valence band and the Fermi velocity
is rather large. In contrast the minority band Fermi surface is due to the As
antisite impurity band and the Fermi velocity is quite small. Therefore, since
in the diffusive limit the conductance is proportional to $v^2$, we expect a
much larger spin polarization of the current compared with the ballistic case.
\begin{figure}[ht]
\epsfxsize=7cm
\centerline{\epsffile{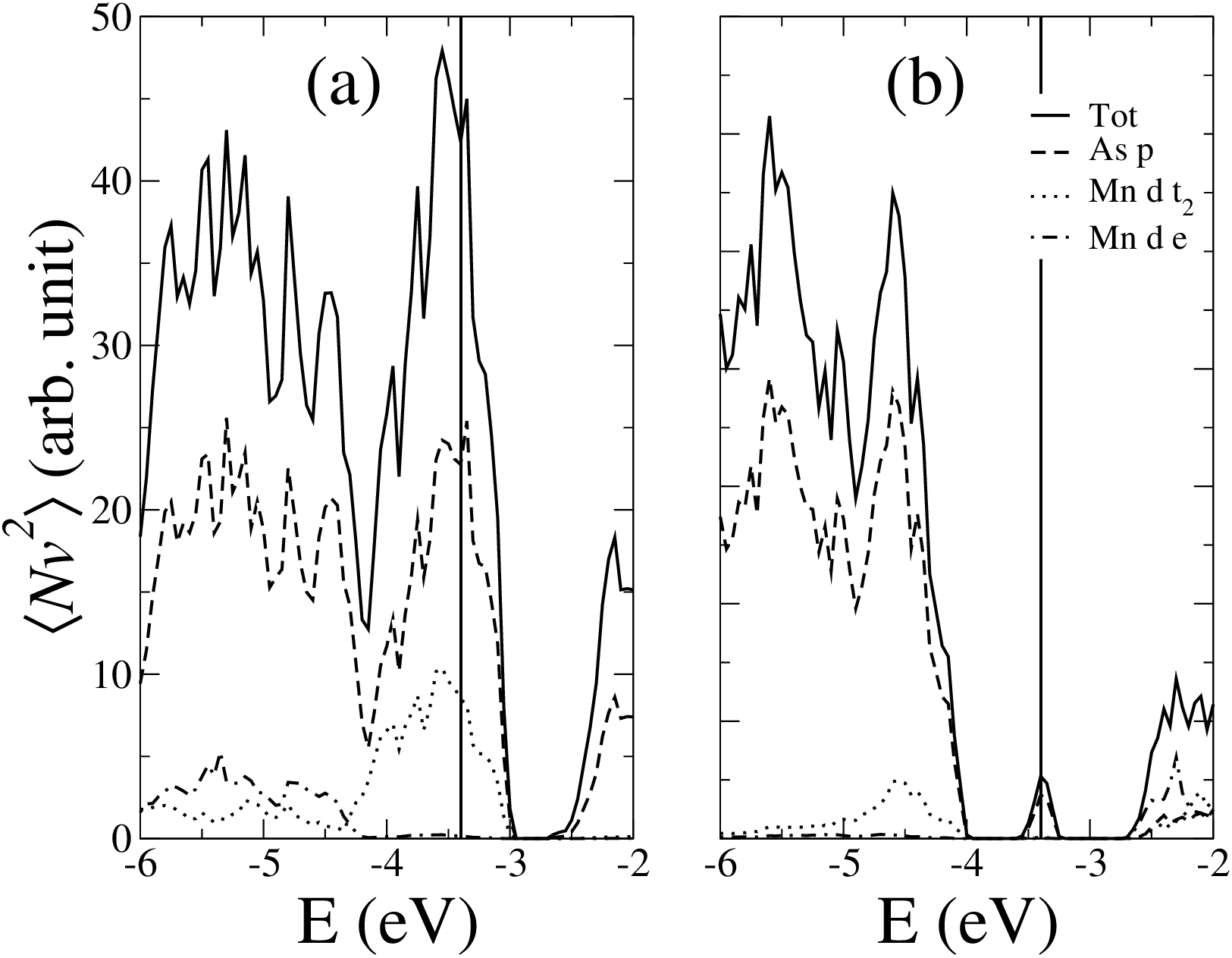}}
%\epsfxsize=6cm
%\centerline{\epsffile{TEST}}
\caption{$\langle N v^2\rangle$ as a function of energy 
for a 50\% Mn DFH with one As$_\mathrm{Ga}$ in the middle of the unit cell, half the way 
between two consecutive (Ga,Mn)As planes. The vertical line indicates the position of the 
Fermi level. (a) majority, (b) minority spins.} 
\label{F9}
\end{figure}

In figure \ref{F9} we present $\langle N {v}^2\rangle$ as a function of energy
for the same DFH as that of figure \ref{F8}. The spin polarization at the Fermi energy
is now 80\%. Since in actual DFH the transport is due to hopping conductance
\cite{Ted}, we can conclude that As antisites, although they destroy the
half-metallic state do not strongly affect the spin-polarization of the current.

\begin{figure}[ht]
\epsfxsize=7.5cm
\centerline{\epsffile{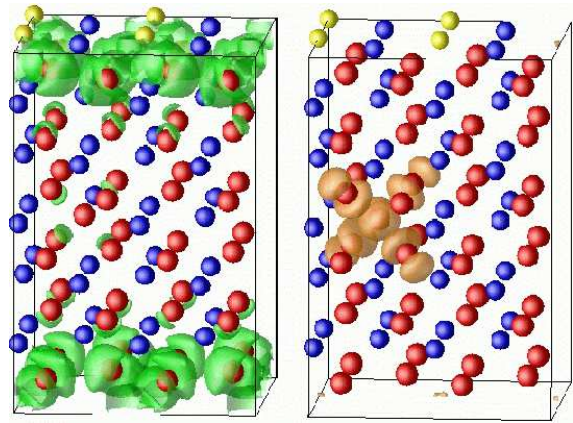}}
%\epsfxsize=6cm
%\centerline{\epsffile{TEST}}
\caption{Real space distribution of the current. This is calculated as 
the charge density distribution in real space of those scattering states 
contributing to the conductance at $E_F$} 
\label{F10}
\end{figure}

Finally it is interesting to study the spatial distribution of the current
across the DFH in the presence of As antisites. In
figure \ref{F10} we present the real space charge density distribution,
$\rho({\bf r})$, calculated only for those states contributing to the 
conductance at the Fermi energy \cite{usprl}. Strictly speaking this does not represent 
the current distribution in real space, but gives
information on the spatial distribution of the conductance electrons'
wave functions around $E_\mathrm{F}$.
The main feature of figure \ref{F10} is that there is a spacial separation 
between the two spin currents, with the majority spin current located near the
Mn plane, and the minority current strongly localized around the As 
antisites. The first is evenly distributed and this is suggestive of a metallic
like behavior, while the second is strongly localized at the scattering center,
suggesting an hopping-like transport. Moreover, if we correlate the spin current
with the relevant band-structure (see figure \ref{F7}), we notice that the 
majority spin-current is hole-like, while the minority is electron-like.

We also investigate how these features change when the As antisite moves toward
the Mn plane. Although the conductance as a function of energy does not present
any significant modifications with respect to the case of figure \ref{F7}, the
spacial distribution shows an increased participation of the As antisite in the
majority spin conductance as it moves closer to the Mn plane. This suggests
that As antisites can play an important r\^ole in spin relaxation processes
within DFH. 

In conclusion, our transport results suggest that, if there is no 
spin mixing, the transport is dominated by a $p$-type metallic-like majority spin 
current with a smaller contribution from an $n$-type hopping-like minority spin current. 
Moreover the two spin-currents are spatially separated, preventing spin-mixing, only if the As
antisites are reasonably far from the Mn plane. 

\subsection{Fit to the Heisenberg Model}

We now investigate the stability of the ferromagnetic state when As antisites
are present. We perform similar calculations to those
described in section IIIC, but this time we include one 
As$_\mathrm{Ga}$ in the unit cell.
Since the relative position of the As$_\mathrm{Ga}$ with respect to the Mn ions is
crucial in determining the electronic and magnetic properties \cite{usapl}, we
investigate how the coupling depends on the As antisite position along the
superlattice direction.

\begin{figure}[ht]
\epsfxsize=7cm
\centerline{\epsffile{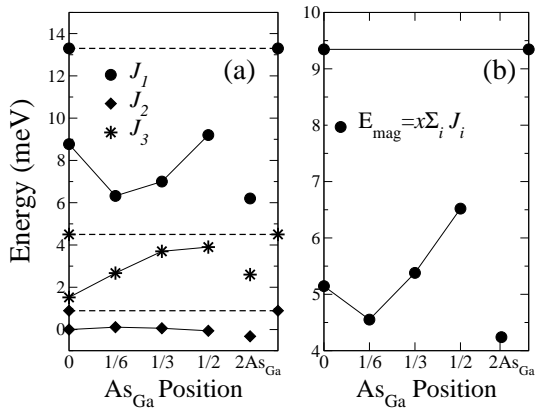}}
%\epsfxsize=6cm
%\centerline{\epsffile{TEST}}
\caption{a) Exchange constants $J_1$, $J_2$ and $J_3$ and b) magnetic energy
$E_\mathrm{mag}$ of a DFH as a function of the position of the As antisite with 
respect to the Mn plane (in unit of the superlattice period). We also include
the case of two As antisites. The horizontal lines indicate the same quantities for the 
antisite-free case.} 
\label{F11}
\end{figure}

In figure \ref{F11}a we present the value of the exchange constant for first,
second and third nearest neighbor couplings as a function of the position of 
the As antisite with respect to the Mn plane. 
In the same figure we also present the same values for the As antisite-free 
case and for the case of two As antisites (located respectively at 1/3 and 
2/3 of the supercell along the superlattice direction). 

From the figure it is clear that, in all cases, most of the coupling comes 
from nearest neighbor interaction, which 
accounts on average for 75\% of the total coupling 
$J_\mathrm{tot}=\sum_iJ_i$. Then there is a fast decay of the exchange
coupling with separation. This coupling is therefore short range. 
Furthermore it is very
interesting to note that the second nearest neighbor coupling (Mn ions at A1
and A2) is almost negligible in all cases, and it sometimes assumes negative
values. This suggests possible local antiferromagnetic coupling between the
Mn ions as recently demonstrated experimentally \cite{mc2}.

Turning our attention to the effect of As antisites, it is 
clear that these weaken the coupling between the Mn ions. This is
expected since an As$_\mathrm{Ga}$ is a donor and therefore its
presence partially compensates the free holes responsible for the long range
ferromagnetic coupling. However our accurate supercell calculations bring
additional interesting features. In agreement with the case of the random
alloys (see reference \cite{usapl}), the actual position of the As antisite with
respect to the Mn ions determines most of the properties. From figure \ref{F11} it is
clear that the ferromagnetic coupling is weakened when the As antisite moves
toward the Mn plane. In particular the exchange constants for the cases when
$z=1/6$ and $z=0$, are very similar to those of the case of two As antisites
($z=1/3$ and $z=2/3$). This feature can be understood by considering the
potential profile discussed in the previous section. In fact the charge
is strongly confined in the Mn plane with a potential barrier separating the
plane from the GaAs region. Therefore it is natural to think that any 
perturbation in the GaAs region will only weakly affect the electronic 
configuration of the Mn plane, unless such a perturbation is spatially located 
in the vicinity of such a plane.

In the second half of figure \ref{F11} we present the magnetic energy, $E_\mathrm{mag}=
x\sum_iJ_{i}$, for the different cells investigated. As we pointed out previously
this quantity is proportional to the Curie temperature, $T_C$.
From the figure it is clear that $E_\mathrm{mag}$ depends sensitively on the 
presence of the As antisites and on their actual location. 
In particular, although we find the lowest value of $E_\mathrm{mag}$ in the case 
in which two As antisites are present, we also find that this is very similar to 
the case of a single As antisite located in close proximity to the Mn plane 
($z=1/6$ in the present case). Therefore we conclude that one can obtain high Curie
temperatures, not only by avoiding the formation of intrinsic defects, but also by
controlling their position with respect to the magnetic region.

Finally we make a few comments on the effect of disorder.
Since in actual samples we are not able to control the exact 
position of the Mn ions with respect to each other, and since the Mn-Mn coupling
is strongly dependent on the relative positions of the Mn ions and those of the As
antisites, it is likely that there are regions of strong Mn-Mn coupling
together with regions of weak or even antiferromagnetic coupling. This suggests
that in addition to configurational disorder, magnetic disorder can also be 
present in DFH even at low temperatures. Therefore since
the electrons (or holes) at the Fermi level have a rather large (metallic) 
density, and are strongly confined in a few atomic planes around the
Mn ions, we can conclude that DFH have the electronic properties of highly 
resistive (dirty) metals. Turning the argument around, we conclude 
that the metallicity is crucial for the magnetic state of DFH, and that the most
metallic samples are likely to show less magnetic disorder and therefore more
robust magnetic properties \cite{McCombe}.

\section{A\lowercase{l}A\lowercase{s}/(G\lowercase{a},M\lowercase{n})A\lowercase{s} heterostructures}

One of the main messages from the analysis done so far is that the exchange part
of the DFT potential creates a strong confinement potential for the majority 
electrons in the Mn plane. In this section we investigate the effects of an 
additional confining potential, namely that obtained 
by sandwiching a (Ga,Mn)As monolayer into the GaAs region of an 
AlAs/GaAs superlattice. Our expectation is that the AlAs/GaAs band 
alignment will further confine the spin-holes in the proximity of the Mn ions, 
therefore enhancing the exchange coupling.

\subsection{GaAs/AlAs band alignment}

\begin{figure}[ht]
\epsfxsize=7cm
\centerline{\epsffile{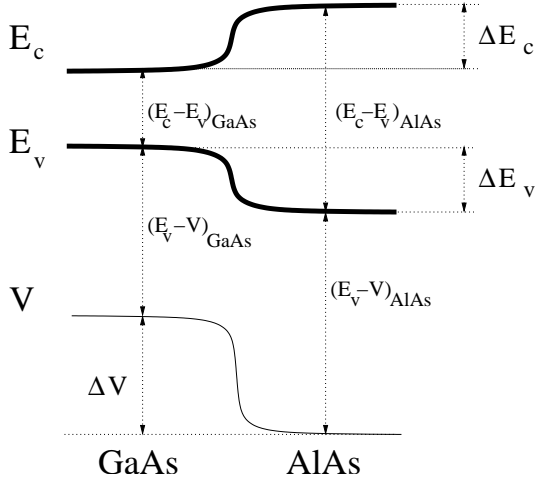}}
%\epsfxsize=6cm
%\centerline{\epsffile{TEST}}
\caption{GaAs/AlAs band alignment.} 
\label{F12}
\end{figure}

Before considering the Mn-doped case we first illustrate the general 
band alignment of an AlAs/GaAs superlattice, as obtained from our DFT 
calculations. The valence band off-set, $\Delta$, is 
calculated as suggested by Baroni et al. \cite{Baroni} as
\begin{equation}
\Delta=\Delta E_v+\Delta{V}\;,
\end{equation}
where $\Delta{V}$ is the off-set between the AlAs and the GaAs electrostatic
potentials calculated for an AlAs/GaAs heterostructure, and
$\Delta E_v$ is the energy difference between the valence band tops, $E_v$, 
of the bulk materials. These are calculated from their electrostatic potential $V$
\begin{equation}
\Delta
E_v=\left(E_v-{V}\right)_\mathrm{GaAs}-\left(E_v-{V}\right)_
\mathrm{AlAs}\;.
\end{equation}
Here we have constructed a supercell by stacking four AlAs cubic cells on top of four GaAs 
cubic cells, all with the same GaAs lattice constant (5.65\AA). Our calculations
give a valence band off-set of $\Delta$=0.405~eV ($\Delta$=0.403~eV if the total
DFT potential is considered instead of the electrostatic one). This value for
$\Delta$ is in very good agreement with both experimental results \cite{galexp} 
and early DFT calculations \cite{galthe}. The resulting GaAs/AlAs band 
alignment is shown in figure \ref{F12}; it provides an additional 
confinement potential for holes in the GaAs region. Therefore a stronger ferromagnetic
coupling is expected.

\subsection{Electronic Structure}

Also in this case the band structure and the DFT potential are the
main quantities to investigate. First we study 
the evolution of the electronic properties as a function of the AlAs
fraction of the superlattice. We construct 
GaAs$_{(5-n)/2}$/AlAs$_n$/GaAs$_{(5-n)/2}$/(Ga,Mn)As$_1$ superlattices, 
where the labels indicate the number of monolayers of the specific semiconductor. 
Note that the total number of monolayers in our supercell is six, and that the AlAs
fraction is always located in the middle of the cell.
\begin{figure}[ht]
\epsfxsize=8cm
\centerline{\epsffile{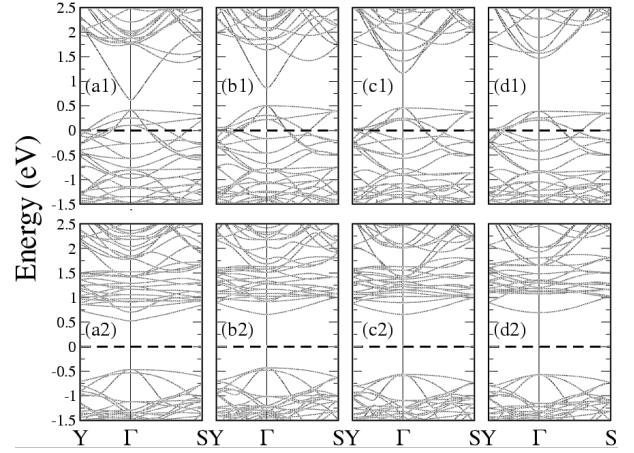}}
%\epsfxsize=6cm
%\centerline{\epsffile{TEST}}
\caption{Band structure of AlAs/GaAs/(Ga,Mn)As DFH as a function of the AlAs fraction: a) $n$=0,
b) $n$=1, c) $n$=3, d) $n$=5. The top panels are for the majority spin band and the bottom for the
minority. The horizontal line indicates the position of the Fermi level, which is set to 
$E_F$=0~eV.} 
\label{F13}
\end{figure}

In figure \ref{F13} we present the band structure of a 50\% DFH with
Mn ions uniformly distributed in the plane (configuration FM$_1$),
for the cases $n$=0,1,3,5. As usual we consider only the in plane
directions.

From figure \ref{F13} we can see that the only appreciable change in the band
structure as the AlAs fraction increases is an enhancement of the band gap. 
This of course reflects the different AlAs/GaAs ratios of the superlattices
and the larger bandgap of AlAs compared with GaAs 
(our LDA values are 1.7~eV and 0.6~eV respectively).
For all the superlattices studied the magnetic moment of the unit cell is
16~$\mu_\mathrm{B}$ and the M\"ulliken analysis gives a Mn $d$ occupation
of $\sim$4.7 and $\sim$0.8 respectively for majority and minority spins. Such an 
occupation is independent of the AlAs fraction and suggests that Mn is always 
in a $d^5$ state plus an antiferromagnetically coupled hole, as in the case of 
GaAs/(Ga,Mn)As DFH. 

A closer look at the band structure reveals another important 
feature. The spin splitting of the valence band top, $\Delta_v=E_v^\uparrow-
E_v^\downarrow$, is a non-monotonic function of the AlAs fraction, with values of
0.88~eV, 0.93~eV, 1.02~eV, and 0.96~eV for $n=$0, 1, 3 and 5 respectively. The
initial increase is due to the enhanced confinement of the free holes in the Mn
region. In fact the mean field expression for the valence band top spin splitting is
simply \cite{Dietl,usprb}
\begin{equation}
\Delta_v=xN\beta\langle S \rangle\;,
\end{equation}
where $N\beta$ is the $p$-$d$ exchange constant, $\langle S \rangle$ is
the mean spin ($\langle S \rangle$=5/2 for (Ga,Mn)As) and $x$ is the Mn 
concentration. The exchange constant, $N\beta$, depends on the degree of
overlap between the hole density and the Mn ions, and this is enhanced by 
hole confinement. Therefore we expect an increase of $\Delta_v$
when the AlAs fraction is increased. The case $n=5$ is different, since
no GaAs region is left and the additional confinement of the hole in the Mn plane
due to the AlAs/GaAs valence band offset is partially lost.

\begin{figure}[ht]
\epsfxsize=7cm
\centerline{\epsffile{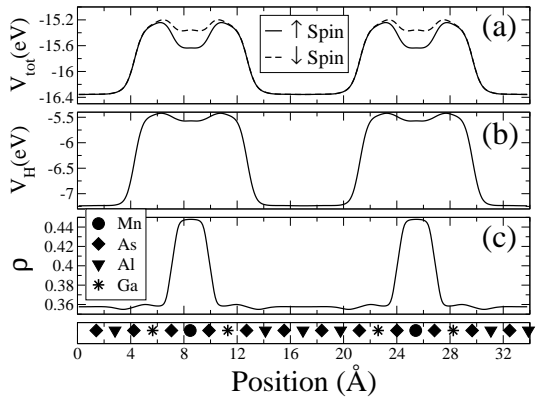}}
\caption{Macroscopic average of a) the total DFT potential, b) the Hartree potential and 
the c) charge density distribution as a function of the position along the superlattice
direction. The system is a GaAs$_{1}$/AlAs$_3$/GaAs$_1$/(Ga,Mn)As$_1$ superlattice
with 50\% Mn uniformely distributed in the plane (configuration FM$_1$). Here
we plot two superlattice periods.} 
\label{F14}
\end{figure}

Another important quantity to investigate is the DFT total potential.
In figure \ref{F14} we present the macroscopic average along the superlattice direction
of the DFT and Hartree potentials, and the electronic charge density,
for a 50\% DFH (configuration FM$_1$), where we introduce three AlAs monolayers
(GaAs$_{1}$/AlAs$_3$/GaAs$_1$/(Ga,Mn)As$_1$).
The figure shows that there is a large well for both majority 
and minority spin electrons in the AlAs region. 
This is mainly due to the Hartree component
of the DFT potential and it is not spin sensitive. The width of this region depends
on the thickness of the AlAs layer and plots for other values of $n$ give similar 
results. If we interpret the macroscopic average of the DFT potential in the 
spirit of the envelope function approximation, we can conclude that new
electrons coming from intrinsic donors will be preferentially localized in the AlAs
region. Therefore the ferromagnetism in AlAs/(Ga,Mn)As DFH appears to be
more robust against electron doping than that in GaAs/(Ga,Mn)As DFH. This is
consistent with the rather large $T_C$ found in AlAs/GaAs/(Ga,Mn)As
DFH \cite{Tanaka,Furdyna}.

\subsection{Fit to the Heisenberg Model}

As in the case of (Ga,Mn)As/GaAs DFH we investigate the strength of the 
ferromagnetic coupling by fitting our total energy calculation to third 
nearest neighbor Heisenberg model. 
We consider only the case of 50\% Mn in the plane and we study
the dependence of the coupling constants on the number of AlAs layers in the
supercell. The results of our fit are shown in table \ref{T3}.
\begin{table} [ht]
\caption{$J_1$, $J_2$, $J_3$ and $x\sum_iJ_{i}$ as a function of the number of
the AlAs layers in the supercell. All the units are meV. \label{T3}}
\centering\begin{tabular}{c|cccc} \hline\hline
AlAs planes ($n$) & $J_1$ & $J_2$ & $J_3$ &  \hspace{0.1in} $x\sum_iJ_{i}$ \\ \hline
$n=0$  & 13.3 & 0.9 & 4.5 & 9.35 \\
$n=2$  & 13.8 & 0.9 & 4.7 & 9.72 \\
$n=3$  & 14.0 & 0.8 & 4.9 & 9.85 \\
$n=5$  & 14.6 & 0.6 & 5.2 & 10.30 \\ \hline\hline
\end{tabular}
\end{table}

From the table one can conclude that the total strength of the coupling, which 
is proportional to $T_C$, increases as a function of the AlAs fraction. This 
is consistent with an enhanced ferromagnetic coupling between the holes and 
the Mn ions due to confinement. Therefore in the absence of intrinsic defects,
AlAs/(Ga,Mn)As DFH are more promising than GaAs/(Ga,Mn)As DFH as high $T_C$ DMS.
It is also interesting to note that the different exchange constants behave in a
different way when the AlAs fraction is increased: $J_1$ and $J_3$ are enhanced
and $J_2$ is reduced. This behavior is not surprising within a carrier mediated
ferromagnetism model (e.g. RKKY), since an increase of the AlAs fraction changes 
the value of the Fermi wave vector, and therefore the period of the exchange 
coupling.

Having established that defect-free AlAs/(Ga,Mn)As DFH present higher $T_C$ than
their GaAs/(Ga,Mn)As counterparts, we finally investigate the stability of the
ferromagnetic coupling against the presence of As antisites. As usual we
introduce one As$_\mathrm{Ga}$ per unit cell at various positions along the
superlattice direction. Here we consider two limiting cases: 1) $n=3$ and the As
antisite is located in one of the GaAs planes adjacent to the Mn plane, 2) $n=5$ and
the As antisite is located in the middle of the unit cell (in the AlAs region).

In the first case, we find the coupling constants to be $J_1=7.0$~meV, 
$J_2=0.0$~meV, $J_3=2.9$~meV, $x\sum_iJ_i=4.95$~meV. 
This is a considerable reduction of the ferromagnetic coupling with respect to 
the defect-free case. In this case the values of the coupling
constants are almost identical to those of GaAs/(Ga,Mn)As DFH in which an
antisite is introduced at the same position (in figure \ref{F11} the 
As$_\mathrm{Ga}$ position is 1/6). Therefore, if the As antisites are introduced
in the GaAs region, there will be no beneficial effects from the AlAs/GaAs band
alignment.

The situation is rather different in the second case where the As antisites are
introduced in the AlAs layer. Now the coupling constants are
$J_1=10.1$~meV, $J_2=-0.1$~meV, $J_3=4.4$~meV, and the reduction of the
total coupling $x\sum_iJ_i$ with respect to the defect-free case is only of 
about 30\%. Interestingly, $J_2$ now assumes a negative value, suggesting some
possible frustration even at low temperature.

In conclusion, AlAs/GaAs/(Ga,Mn)As DFH have stronger ferromagnetic
interaction between the Mn ions than GaAs/(Ga,Mn)As DFH. Moreover, in the case
that intrinsic defects are kept into the AlAs region, the ferromagnetic order is more
robust against hole compensation. 

\section{Conclusions}

We have performed an extensive theoretical study of the electronic, magnetic and
transport properties of GaAs/(Ga,Mn)As and AlAs/GaAs/(Ga,Mn)As DFH, using 
DFT within LSDA. 

We find that GaAs/(Ga,Mn)As DFH show an half-metal band structure with metallic 
conductance in the Mn plane. The macroscopic average of the DFT potential indicates
a selective confinement of the spin holes in the Mn planes and the Mn-Mn ferromagnetic
interaction is non-monotonically dependent on the Mn concentration.

When compensating defects such as As antisites are introduced, the half-metallic
state is lost and conducting channels appear in the minority spin band. These are
due to hopping conductance through localized As$_\mathrm{Ga}$ states. However,
at least when the As antisites are far from the Mn planes, there is a spatial 
separation of the two spin currents with a metallic majority spin current located 
in the Mn planes, and an hopping-type minority spin current located primarily in the
GaAs region. These differences in the type of transport for the two spin bands 
are magnified in the diffusive limit, for which we calculate a spin polarization of
about 80\%. Finally, the presence of As antisites generally weakens the ferromagnetic 
interaction, and local antiferromagnetic coupling between Mn ions is possible
at low temperature.

With all these results in hand we conclude that GaAs/(Ga,Mn)As DFH behave as
dirty planar metals, where the strength of the ferromagnetic coupling depends
strongly on the amount and the position of the intrinsic defects.

Finally we have investigated the effect of additional confinement by
studying AlAs/GaAs/(Ga,Mn)As DFH. In this case our accurate total energy
calculations confirm that the band offset between
GaAs and AlAs strongly confines the holes in the Mn region resulting in a larger
Mn-Mn coupling. In addition we find that these structures are less sensitive to the 
presence of As antisites than the simpler GaAs/(Ga,Mn)As DFH.

We therefore conclude that band engineering, selective doping and the ability
to control the actual position of As antisites or other compensating defects
are the main keys to obtain room temperature ferromagnetism in DFH. 

\section*{Acknowledgments}
This work is supported by Enterprise Ireland under the grant EI-Sc/2002-10.
The calculations have been carried out using the computational facilities of the
Trinity Centre for High Performance Computing (TCHPC).
Useful discussions with N.A.~Spaldin, and J.M.D.~Coey are kindly 
acknowledged.

\end{document}